# Advancing Mixed Reality Game Development: An Evaluation of a Visual Game Analytics Tool in Action-Adventure and FPS Genres

PARISA SARGOLZAEI, Faculty of Business and Information Technology, Ontario Tech University, Oshawa, Ontario, Canada

MUDIT RASTOGI, Computer Science and Technology, SRM Institute of Science and Technology, Kattankulathur, Tamil Nadu, India

LOUTFOUZ ZAMAN, Faculty of Business and Information Technology, Ontario Tech University, Oshawa, Ontario, Canada

In response to the unique challenges of Mixed Reality (MR) game development, we developed *GAMR*, an analytics tool specifically designed for MR games. *GAMR* aims to assist developers in identifying and resolving gameplay issues effectively. It features reconstructed gameplay sessions, heatmaps for data visualization, a comprehensive annotation system, and advanced tracking for hands, camera, input, and audio, providing in-depth insights for nuanced game analysis.

To evaluate *GAMR*'s effectiveness, we conducted an experimental study with game development students across two game genres: action-adventure and first-person shooter (FPS). The participants used *GAMR* and provided feedback on its utility. The results showed a significant positive impact of *GAMR* in both genres, particularly in action-adventure games. This study demonstrates *GAMR*'s effectiveness in MR game development and suggests its potential to influence future MR game analytics, addressing the specific needs of developers in this evolving area.

CCS Concepts: • Applied computing~Computers in other domains~Personal computers and PC applications~Computer games • Human-centered computing~Human computer interaction (HCI)~Interaction paradigms~Mixed / augmented reality • Human-centered computing~Visualization~Visualization application domains~Visual analytics

This work is supported by NSERC Discovery and Mitacs Globalink Research Internship.
Author's addresses: Parisa Sargolzaei and Loutfouz Zaman, Faculty of Business and Information Technology, Ontario Tech University, 2000 Simcoe St N, Oshawa, ON L1G 0C5, Canada; Mudit Rastogi, Computer Science and Technology, SRM Institute of Science and Technology, SRM Nagar, Kattankulathur, Chengalpattu District, Tamil Nadu 603203, India.

## 1 Introduction

The rapid growth and advancements in Mixed Reality (MR) technologies present a unique opportunity to develop game analytics tools specifically designed for MR environments. MR, distinct from traditional gaming platforms, merges real and virtual elements, creating a novel gaming experience that necessitates a different approach to analytics. This integration of digital and physical worlds in MR games adds complexity to player interactions.

Mixed Reality (MR) devices face several hardware limitations, including a narrow field of view [44] and imperfect hand tracking [15], both of which are crucial for an optimal MR experience. To ensure the best possible gameplay and user experience, game developers and designers must carefully adjust and address these limitations. Therefore, it is crucial to analyze gameplay data within the MR environment, leveraging the capabilities of MR devices to gain deeper, more contextual insights into player behavior and game dynamics. Such insights can lead to more engaging and immersive gaming experiences.

MR's highly immersive nature is key to understanding player behavior in a more natural gaming environment. For instance, the immersive experience of *Pokémon Go* is a testament to this in Augmented Reality (AR) gaming [37]. Shin [41] emphasizes the importance of technological and affective affordances in enhancing AR game immersion. Tailored analytics for AR/MR can effectively measure the impact of this immersion. MR devices enable real-time analysis of player behavior in their physical surroundings, offering nuanced insights into how players navigate and respond to stimuli. This integration, enhancing both immersion and engagement, opens new avenues for gameplay analysis, particularly in understanding player interactions within their physical environment. The spatial nature of MR environments is crucial for insights into player interactions in 3D space, facilitating a more natural interaction with data through the placement and manipulation of digital objects in real space.

From a Games User Research (GUR) perspective, the immersive experience provided by MR devices allows for intuitive exploration and manipulation of data in 3D space, enhancing the understanding of complex datasets. This approach, known as *immersive analysis*, has been recognized for its potential [8, 25, 31]. MR offers new types of data, such as player movement in physical space, interactions with virtual objects, and gaze tracking, providing insights into player behavior and preferences beyond what traditional PC-based game analysis can offer. The significance of new data types has been previously highlighted in the context of wearable technology by Kim and Shin [22].

MR's multimodal nature, encompassing visual, auditory, and haptic elements, enables a comprehensive understanding of its impact on the gaming experience. Analyzing MR games can provide insights into innovative game design elements that merge physical and digital worlds, potentially leading to new gaming genres. Additionally, the multimodality of MR can be used to assess the physical and cognitive load on players during gameplay, offering a more holistic view of the gaming experience. In a related context, technologies like *Kinect* have been successfully used to assess kinematic strategies of postural control [10].

Despite significant progress in immersive analytics, many studies have highlighted a major shortfall in handling complex or precise data inputs, which impedes effective visual analysis in MR environments [48].

The existing landscape of game analytic tools predominantly caters to PC games, leaving a gap for aspiring developers working on mixed reality (MR) games. Consequently, these developers may find themselves resorting to labor-intensive and time-consuming practices, such as manually reviewing lengthy video recordings of player game sessions, to identify and address bugs and issues within their games. We present *GAMR* (Game Analytics – Mixed Reality), our MR game analytics tool, which aims to fill the gap in game analytics for MR games by offering a comprehensive toolset for analyzing and evaluating player behavior. Our work makes the following contributions:

- **Creation of a Specialized MR Analytics Tool**: We have developed *GAMR* with comprehensive analytical capabilities, including reconstructed gameplay sessions, heatmaps for data visualization, an annotation system, and tracking features for hands, camera, input, and audio. This specialized tool is tailored to meet the unique demands of MR game analytics.
- **Development of Two Genre-Specific Games for System Evaluation**: To assess the effectiveness of *GAMR*, we have created two games, one in the Action-Adventure genre and the other a First-Person Shooter. These games serve as platforms for evaluating the system's capabilities in different gaming contexts.
- **Mixed Methods User Studies for Evaluation**: We conducted two mixed-methods user studies focusing on the usability and usefulness of *GAMR*. These studies provide a comprehensive evaluation of the system from the perspectives of end-users, offering valuable insights into its practical application and impact.

## 2 Related Work

### 2.1 Game Analytics

Medler et al. [32] developed *Data Cracker* for *Dead Space 2* to monitor gameplay and increase data literacy of the development team. The tool uses telemetric "hooks" within the game's code to gather player data, which is then visually aggregated. *GAMR* records player data independently from the code, which simplifies integration. Stahlke et al. [43] developed *PathOS*, a tool simulating user testing through agents mimicking player navigation. The purpose is to simplify development and reduce the need for extensive playtesting. *PathOS* incorporates a model of player perception, memory, and cognitive architecture to simulate visual and information processing. By leveraging three visualizations – individual path and bubble visualizations, as well as heatmaps – users can determine the best travel routes in the game. Similarly, *GAMR* features individual path visualization and heatmap, which shows the collective movements of all players.

Dixit and Youngblood [11] presented *PlayerVis*, a tool to comprehend player behavior in 3D environments using visual data mining and processing of logged player data. The tool visualizes player trajectories in a simplified version of a game world to analyze behavior and identify key patterns. Similarly, *GAMR* displays player paths in the real-world game space, which helps users to develop levels adaptable to different player behaviors, instead of the ideal player. MacCormick et al. [26] presented *FRVRIT* for VR game analysis, which tracks full-body player interactions and uses voxel heatmaps to show movement data. *FRVRIT* enables the user to analyze body movements directly using VR's dynamic camera. Similarly, *GAMR* uses the camera features of *HoloLens 2*, which allows data visualization from multiple perspectives.

MacCormick et al. developed and evaluated *Echo* [27], a tool for analyzing player behavior by reconstructing gameplay from recorded data and allowing to replay it from different perspectives. Its expanded version, *Echo+* [28], was evaluated on four different genres of games. *GAMR's* borrows the ideas for its replay system from *Echo/Echo+*.

Andersen et al. [5] developed *Playtracer*, which uses a graph to represent game states as nodes and player paths as edges, adjusting node sizes based on the number of players reaching each state. Similarly, *GAMR* visualizes data from multiple players and allows analysis of player paths, interactions, and transitions. Ahmad et al. [2] developed the Interactive Behavior Analytics (IBA) to model players' behavior. The tool uses abstraction algorithms for modeling and visualization, and clustering algorithms for labeling to categorize behaviors, providing an in-depth insight into player actions. Both components are designed to analyze and model player actions, and to provide insights for the user. In contrast, *GAMR* takes a unique approach by recording diverse game metrics for modeling and examination.

Charleer et al. [9] argue for the importance of game metrics in understanding player behavior. *GAMR* extracts these metrics from game sessions, which allows the user to look closer into gameplay insights. Drachen and Canossa [12] also discussed the role of gameplay metrics in player testing. Drachen and Canossa conducted two studies on visual representations in commercial games to analyze player behavior. The first aimed to assess level design and displayed player death locations on a 2D game level. The second compared developers' intended path with actual player paths. Similarly, *GAMR* aims to preserve game context. By reconstructing gameplay sessions, *GAMR* aims to provide a comprehensive understanding of player behavior while retaining the contextual information of the original game environment. The preservation of game context within *GAMR* enhances the analysis and interpretation of player interactions, enabling game developers and designers to make informed decisions for game improvements and adjustments.

Emmerich and Masuch [16] delve into the significance gameplay metrics in analyzing player behavior, underscoring the need to capture player interactions and social dynamics in games. Emmerich and Masuch introduce three social metrics: social presence, player cooperation, and leadership. In the context of *GAMR*, the inclusion of gameplay metrics aligns with the broader goal of understanding player behavior and interactions, providing valuable insights for enhancing social dynamics and engagement in MR games. Osborn et al. [35] developed *Gamalyzer*, a tool for tracking players' decisions and actions in the game. Each play trace is shown with a vertical line where key game events are marked. Similarly, *GAMR* records player actions, such as player input. In *GAMR*, users can use the input tracker to observe players' actions. Users then can examine the specific details of triggered events and their outcomes.

Drenikow et al. [14] developed a tool that simulates players' behavior and measures arousal states. By tracking both players' movements and physiological measures, the tool provides a more comprehensive understanding of player experience. This can allow users to analyze physical interactions, as well as emotional and cognitive aspects that may contribute to player engagement. In *GAMR*, tracking player movements is also present.

Wallner and Kriglstein [45] developed *PLATO*, a visual analytics system for multidimensional gameplay data. *PLATO* uses path finding, clustering, and diverse visualizations to get insights on gameplay behavior. Similarly, in *GAMR*, the integration of data visualization with data analysis is one of the main features. By combining these two aspects, *GAMR* works toward facilitating exploration and understanding of gameplay data. Nguyen et al. developed *Glyph* [34], a tool which allows individual and collective analysis in puzzle games. *Glyph* features an aggregated data overview and visualizes individual player behavior. Similarly, *GAMR* supports multi-data visualizations, enabling users to explore and analyze player behavior and interactions from both individual and collective perspectives.

Kriglstein et al. [24] studied the impact of data visualizations, particularly heatmaps, in analyzing and visualizing games. Heatmaps were used to provide an overview of the relative data densities, which focused on the usage of melee weapons in a team-based FPS game. Similarly,

*GAMR* incorporates heatmap visualization as well. Heatmaps in *GAMR* visualize player movement densities and help identifying hotspots, patterns, and areas of interest.

## 2.2 Spatio-temporal Analytics

Kang and Kim [20] examined the topic of spatio-temporal analysis, which encompasses visualization approaches, trajectories, and an in-house telemetry system. The goal of this work was to facilitate understanding of game data. Kang and Kim highlighted the importance of spatio-temporal analysis, which is one of the main features of *GAMR*. Kang and Kim proposed a visual analysis technique that simplifies data analysis. Using spatio-temporal data, Kang and Kim propose how to visualize player behavior in the game world directly, which can ultimately improve the game development process. The integration of spatio-temporal analysis in *GAMR* aligns with Kang and Kim's findings and underscores the importance of understanding player behavior within the context of time and space. Drachen and Schubert [13] stressed the importance of spatial and spatio-temporal game analytics in game user research and the gaming industry. Schubert particularly calls attention to the use of heatmaps for visualization. Within *GAMR*, heatmaps visualize players' movement using point data (*X*, *Z* coordinates), which helps to understand how level design affects player behavior. Wallner et al. [46] created a tool similar to *GAMR* that records spatial game information. The tool visualizes the game space using nodes representing areas players traverse. This visualization offers a holistic view of player movement and interactions, enabling analysts to discern player behavior, detect patterns, and evaluate game design and flow. Kloiber et al. [23] developed a VR system for visualizing human motion data, analyzing user behavior through motion paths and hand movements. Similarly, *GAMR* employs hand motion tracking, allowing users to study player movement and hand actions within the game.

## 2.3 Temporal Analytics

Kim et al. developed *TRUE* [21], a framework for capturing analytics in intricate systems. It prioritizes recording events with timestamps and integrates questionnaires, allowing players to offer attitudinal feedback. Thus, *TRUE* provides a holistic method for data capture and analysis in complex systems. *TRUE* and *GAMR* both work toward comprehensive data capture and analysis. However, while *GAMR* does this through metrics and playback, *TRUE* focuses on recording timestamped events and using questionnaires for player feedback. Feitosa et al. developed *GameVis* [17], a framework for incorporating game data visualization into web technologies, aiding developers in crafting tailored analytics visualizations. The framework underscores displaying data temporally and emphasizes the importance of time-based representation similar to *GAMR*.

## 2.4 Mixed Reality Analytics

Sicat et al. [42] introduced *DXR*, a *Unity* toolkit for extended reality (XR) applications in the Unity engine. *DXR* allows users to craft immersive 3D data visualizations swiftly without needing expertise in 3D graphics, streamlining the development of XR prototypes. Similarly, *GAMR* offers immersive and interactive data visualizations to assist users in developing MR games. Büschel et al. [7] developed *MIRIA*, a toolkit specifically designed for analyzing and in-situ visualizing user interactions within MR environments. This toolkit offers a range of visualizations, including heatmaps, 3D trajectories, and scatterplots, to facilitate comprehensive data analysis. *MIRIA* primarily focuses on spatial data analysis in AR and MR applications, rather than games. However, it is worth noting that in *GAMR*, in-situ data visualizations for MR games are also a prominent feature, which provide insights into player interactions. Reipschläger et al. [38] introduced *AvatAR*, a tool for visualizing human motion using 3D trajectories and virtual avatars, providing a detailed representation of movement and posture. Similarly, *GAMR* captures and visualizes hand gestures within gaming environments, deepening the understanding of player behaviors during gameplay. Hubenschmid et al. [19] introduced *ReLive*, a tool for analyzing MR user studies by interacting with spatial recorded data within the original environment. Similarly, *GAMR* lets users engage with recorded game data, such as audios and inputs, in the exact context of the original gameplay. Alexiadis et al. [3] created a *Kinect*-based tool for evaluating dancers' performances in a 3D virtual space using recorded human skeleton data. It assesses based on joint positions, velocities, and 3D flow error scores. Similarly, *GAMR* tracks hand joints for in-depth visualization and analysis of player behavior within the gaming setting.

## 2.5 Comparison with Existing Tools

Existing tools can track player's HMD position and orientation but often overlook crucial factors like controller movements, hand positions, and audio feedback, essential for player immersion and comfort, particularly in MR devices like *HoloLens* that heavily rely on hand-gesture interactions [49]. Recent studies highlight the complexity of hand-gesture interactions compared to traditional input devices [33]. *GAMR* captures these additional data points to provide developers and designers with a deeper understanding of player experience and interactions, particularly in games where precise gestures are essential, thereby enhancing gameplay and user satisfaction. Table 1 contrasts the features and limitations of *GAMR* relative to other tools mentioned in the subsections above.

Table 1. Highlighting the key features and limitations of *GAMR* in comparison to other related tools.

| Tool/Framework | DXR | MIRIA | AvatarAR | Echo+ | GAMR |
|---|---|---|---|---|---|
| Primary Focus/Domain | XR | MR | MR/AR | PC | MR/AR |
| HMD Movement Tracker | | ✓ | ✓ | | ✓ |
| Full Body Movement and Posture Tracker | | | ✓ | | |
| User Interaction Analysis | | | | | ✓ |
| Custom Object Tracking | | | | ✓ | ✓ |
| Game Data Analytics | | | | ✓ | ✓ |
| Audio Tracker | | | | | ✓ |
| 2D/3D Data Visualizations | ✓ | ✓ | | | ✓ |

## 3 *GAMR*

*GAMR*[2], designed for MR games, is an analytic tool that reconstructs game sessions from recorded gameplay data and operates as a *Unity* [50] plugin. Tailored for the *HoloLens 2* [51] MR headset, it aids developers in understanding player behaviors within the MR environment. *GAMR* consists of two components: recording and visualization, and features gameplay session reconstruction (replay system), heatmap visualization, and an annotation system. These features were inspired by other analytic tools, keeping the nuances of MR games in mind.

*GAMR* was developed using the agile development approach. Developing *HoloLens* applications requires prioritizing user experience and interface design due to the immersive nature of AR. Traditional requirements engineering methods struggle to incorporate users' surroundings effectively, making iterative processes essential for optimizing data presentation in 3D space. Agile approaches, with their flexibility, adapt swiftly to new insights and technological advancements.

---

[2] https://github.com/parisasrg/GAMR

During *GAMR* development, we involved users in testing to gather feedback on its UX/UI and identify feature improvements before starting phase 2. Their feedback, coming from individuals with experience in both game development and game analytics tools, ensured that the analytic tool met our goals. For instance, testers suggested that in the playback system, especially when analyzing multiple sessions, navigation would be easier if the line showing explored areas extended progressively as the player character moved, rather than displaying the entire explored area at once. They also suggested that the playback should automatically pause when using the annotation system to prevent missing any details while taking notes. Continuous user feedback and exploratory design enhance AR game analytics, fostering creativity and innovation. Flexible design methods address visual and spatial challenges, efficiently allocating resources during prototyping and user testing stages. This approach has been validated through successful case studies (see e.g., [30]) and our internal iterative testing.

## 3.1 Recording System

In video games, game objects are primarily categorized into dynamic and static types [40]. Dynamic objects, including characters, enemies, and moving platforms, change their position and rotation throughout gameplay. In contrast, static objects, such as walls and buildings, remain stationary, forming the game's backdrop. *GAMR*'s system captures and logs (into a text file) details about both these object types, including their positions and orientations.

Non-player characters (NPCs) are essential components in most game genres, and their movement during gameplay can greatly influence the player experience [4]. Current game analytics tools for VR and MR games are unable to track NPCs, including enemies, player characters, and user-defined custom objects, limiting their ability to provide insights into navigation issues and challenges. To address this gap, *GAMR* allows users to add and track an unlimited number of dynamic game objects, providing greater flexibility and comprehensive data collection.

## 3.2 Recording System

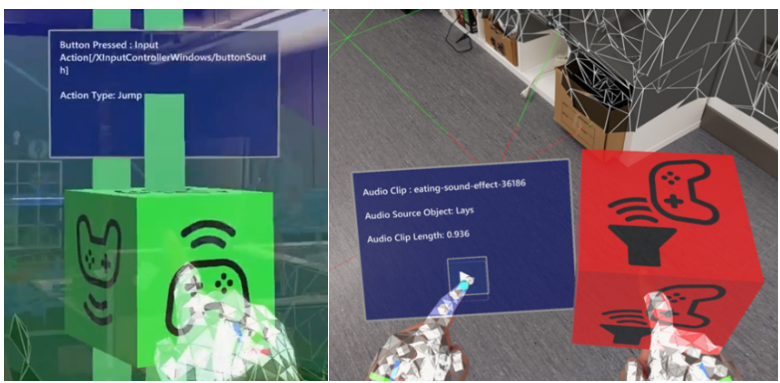

Fig 1. A showcase of input tracker (left) and audio tracker (right) in *GAMR*.

Within *GAMR*, various game objects can be tracked using the recording system. While some objects have specific tracking systems tailored to their unique properties, others have their general information, such as position and rotation, tracked. The following are game objects that can be tracked within *GAMR*:

1) **Main Camera:** The main camera, which represents the viewpoint of the *HoloLens 2* headset, is tracked in terms of its position and orientation. Additionally, the view frustum, which defines the region visible to the camera, is also recorded.
2) **Player:** The player character, representing the user in the game, can be tracked in terms of its position and orientation.
3) **Objects' Audio:** *GAMR* captures game audio data, including positional audio cues, audio names, lengths, and their linked audio source game objects (see Fig 1). This feature provides precise tracking of audio playback and timing within the game, a critical aspect in MR and AR game development for enhancing immersion and gameplay [39].
4) **Player Input:** *GAMR* records traditional gamepad controls and *HoloLens 2* hand gestures. It logs button presses, joystick movements, trigger activations, and hand gestures like grabbing, pointing, or swiping, associating them with their positions in the game world. These recorded inputs displayed as 3D square boxes, offering users the capability to interact with them to access associated information, including the button's name and the player's corresponding in-game action (see Fig 1). Additionally, with the hand tracker, users gain insight into the real-world position and movement of player hands, accurately reflecting their actions within the context of the game session.
5) **Customizable Game Objects:** Users can select specific game objects for tracking. The tool monitors these objects' positions and orientations during the game session.

*GAMR* is versatile, adapting to various game environments and objects. While it captures player and camera data by default, it can also record custom game objects. This feature enables deeper analysis of object interactions within the game, offering insights into gameplay and player experience. For instance, tracking interactive objects like puzzle pieces reveals player interactions and their gameplay impact.

### 3.3 During Gameplay

In *GAMR*, recording can be started via the "Record" button in the playback menu or activated at the game's start through the *Unity* plugin.

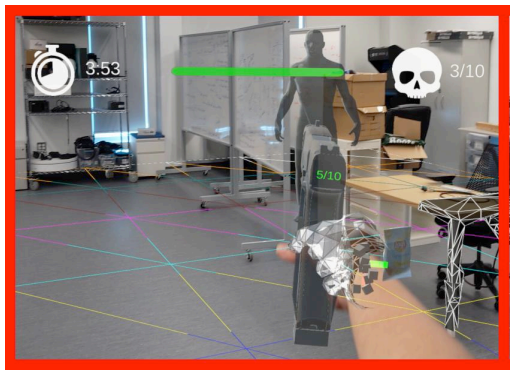

Fig 2. An example of recording tracked objects including hand gestures and weapon object movement.

When the recording begins, the system identifies and registers all trackable objects in the scene, including players, cameras, and other dynamic game elements (Fig 2). Dynamic objects are recorded continuously during the game session, while static objects have data captured once the saving process starts.

Throughout gameplay, data points for dynamic objects are continuously added by the recording system. When the saving process is initiated, either manually or automatically, these data points are sent to the saving system. Objects are recorded at set time intervals from the start to end of recording, ensuring synchronized data capture for each object based on when events occurred.

### 3.4 *GAMR*'s UI Manager

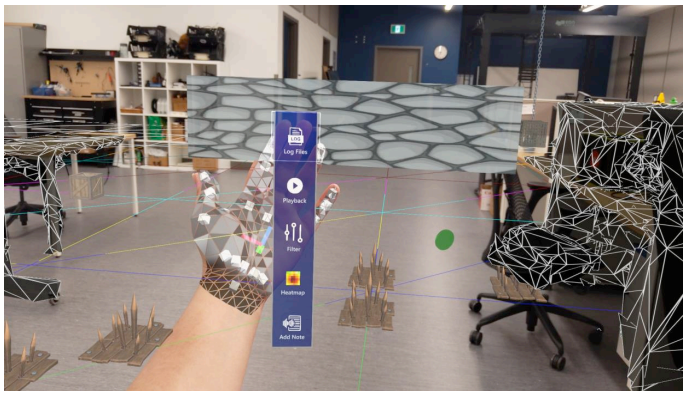

Fig 3. *GAMR*'s hand menu and UI manager.

In *GAMR*, users can access menu options by looking at their palm while wearing the *HoloLens 2*, allowing intuitive interactions without external devices. The menu options available in *GAMR*'s hand menu include:

1) **Log Files**: *GAMR* auto-detects log files post-recording, populating a list for easy selection, eliminating manual searches.

2) **Playback**: Users have full control over recording and replay, including fast-forward, rewind, pause, and resume functions.

3) **Filter**: Users can toggle visibility of recorded objects to customize their analysis based on research interests.

4) **Heatmap**: Displays a 2D AR projection of player movements in the game space, color-coded by density.

5) **Annotation System**: Users can voice-record annotations, which are auto-converted to text using speech recognition.

Upon pressing the "Load" button in *GAMR*'s playback menu, tracked objects from the recorded session populate the scene. Users can toggle specific objects on or off, focusing on game elements of interest. They can also visualize players' movement trajectories throughout the session, either as an overview or in real-time replay. This offers insights into level design efficiency, player strategies, and gameplay experiences.

### 3.5 *GAMR*'s Features

*GAMR* offers three key features to aid game developers and designers. Their functionalities and benefits are detailed in the subsequent sections. *GAMR*'s user interface (UI) menu grants the user complete control over these features.

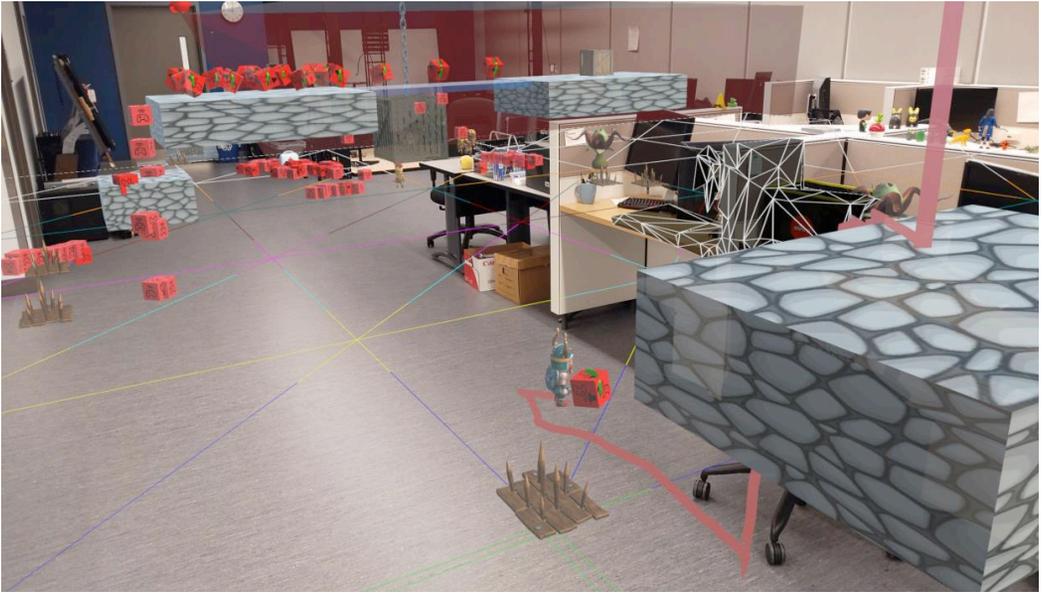

Fig 4. *GAMR*'s tracked objects in the Action-Adventure game.

#### 3.5.1 Gameplay Session Reconstruction (Playback System)

The first and most important feature of *GAMR* is its playback system. *GAMR*'s recording system and playback menu enable users to reconstruct the original gameplay session using the recorded game metrics and data. When the recording system is active, all tracked objects, including players and other dynamic elements, have their positions stored and saved into log files (see Fig 4). When the playback is initiated, these objects are placed back in their initial positions at the start of the recording. To support multi-data analysis, each recorded player's data points are displayed with a unique color, distinguishing them from other recorded players (see Fig 5). This allows for easy identification and analysis of individual player behaviors and interactions.

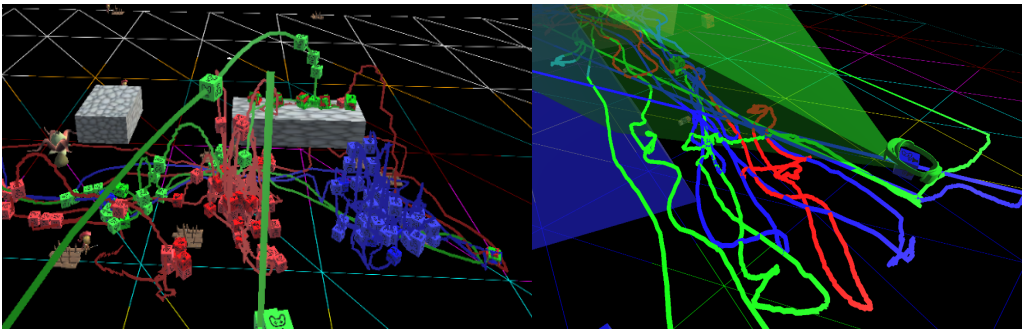

Fig 5. Color mapped players' data in *GAMR*.

Pressing the "Play" button in the playback menu initiates movement of dynamic objects based on the recorded timeline. This allows users to view the replayed gameplay session, analyzing player actions and identifying potential areas for improvement.

#### 3.5.2 2D Aggregated Data Visualization (Heatmap)

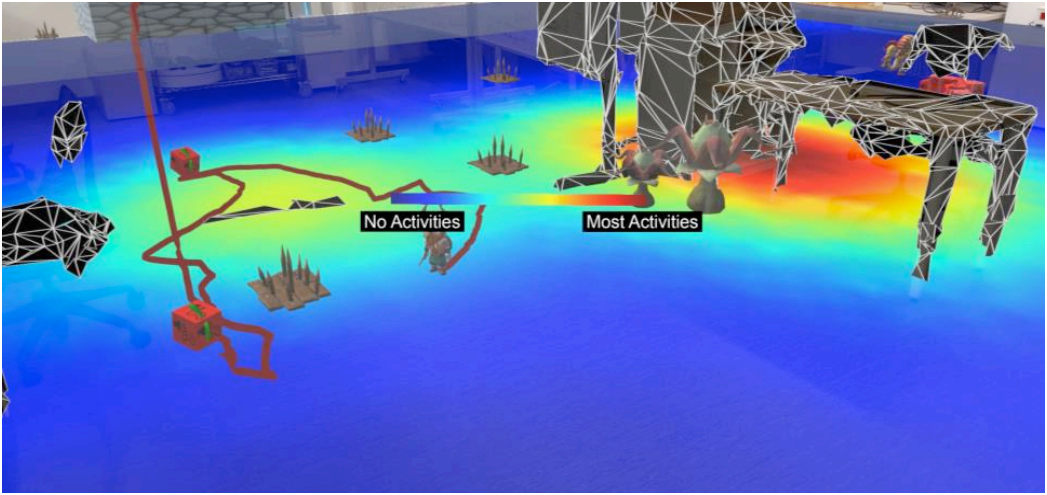

Fig 6. *GAMR*'s heatmap depicting players' movements, with areas of high activity represented in red and areas of low activity shown in blue.

*GAMR* generates a 2D heatmap using *X* and *Z* coordinates from 3D vectors to visualize player movements within MR environments. The feature was designed to aid developers in understanding player navigation and interaction. The heatmap (Fig 6) uses a color scheme, with red highlighting areas frequented by players and blue for less visited areas [13]. The colors range from blue (coldest) to red (warmest). Using the heatmap, users can spot underutilized game areas and identify intended player routes ("golden paths"). By analyzing high-density player activity areas, they can optimize paths aligning with players' preferred routes, guiding them toward specific objectives.

### 3.5.3 Annotation System

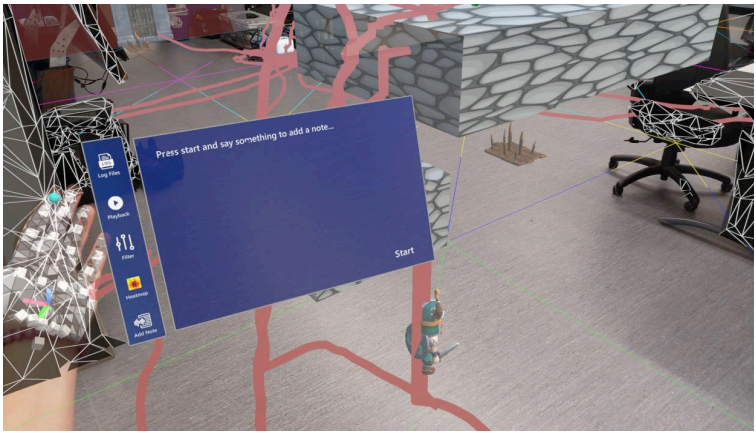

Fig 7. *GAMR*'s annotation system to record notes during analysis using Speech-to-Text feature in *HoloLens 2*.

*GAMR* includes an annotation system (Fig 7), allowing GUR researchers and game developers to document observations during playtesting. Using the *HoloLens 2*'s speech-to-text feature, spoken words are converted into text notes. This eliminates manual notetaking, letting users concentrate on gameplay and observations. Upon saving a note, an icon representing the annotation will automatically appear at the corresponding location. When a note is saved, an annotation icon appears at the relevant location. Users can click this icon to view observation details without removing the *HoloLens 2*, facilitating later analysis.

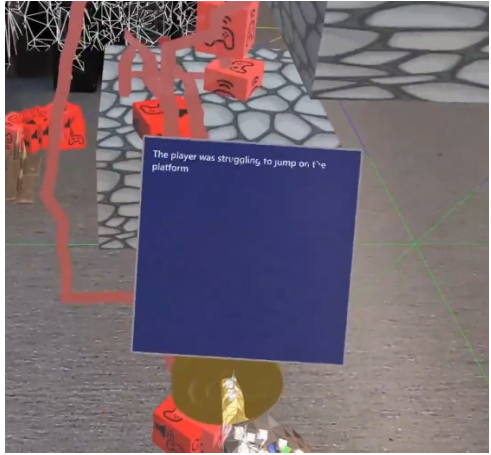

Fig 8. A presentation of the exhibited note resulting from the user's interaction with the generated note object.

#### 3.5.4 Camera and Hand Tracking

Additionally, *GAMR* offers two other features that can greatly assist users in their game analysis and visualization.

*GAMR* includes a camera tracker feature, allowing users to understand player movement and exploration in the game environment, capturing the player's field of view. This helps identify unexplored areas and potential object placements. It supports both first-person, highlighting the player's camera movements, and third-person perspectives, showcasing the controlled character and the human player's movements and view.

*GAMR* also features hand tracking, useful for games utilizing hand gestures. It identifies potential detection issues with the *HoloLens 2* by recording joint movements and wrist positions. When loaded, it updates based on recorded data, enabling users to assess hand movement accuracy in gameplay and pinpoint potential improvements.

## 4 Games

We developed two games in FPS and Action-Adventure genres due to the current unavailability of open-source MR projects for *HoloLens 2*. These games, tailored to our lab's layout, were created to assess *GAMR*'s effectiveness and performance.

By evaluating *GAMR* with an FPS and an Action-Adventure game, we can understand diverse player behaviors and interactions. Different genres bring unique design challenges, and by using both, we can refine *GAMR* for broader applicability. The following sections delve into the details of these two games.

### 4.1 Action-Adventure Game

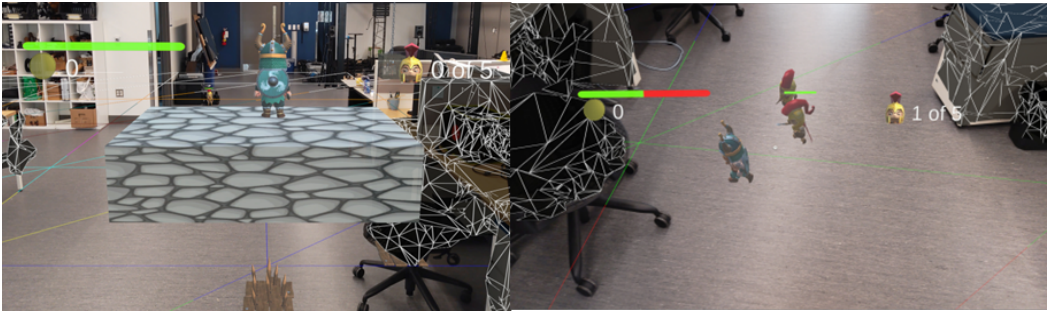

Fig 9. A preview of the Action-Adventure game.

Our first game for *GAMR* analysis is an action-adventure inspired by *Astro Bot: Rescue Mission* [52], designed for the *HoloLens 2* headset. Spanning three levels, it integrates platform elements and is controlled via an *Xbox* controller. The game utilizes *HoloLens 2*'s spatial mapping for real-world object interaction.

In this third-person game, players adopt the perspective of a mini-Viking character who embarks on battles against different enemies (see Fig 9). Actions such as attacking and jumping are executed using corresponding buttons on an *Xbox* controller. The primary objective is to defeat enemies and accumulate points by collecting scattered coins throughout each level. Utilizing the layout mapped by *HoloLens 2*, players can acquire collectibles and evade enemies. Each level presents its own set of challenges, featuring various enemy types, obstacles such as spikes and lethal plants, and engaging boss fights.

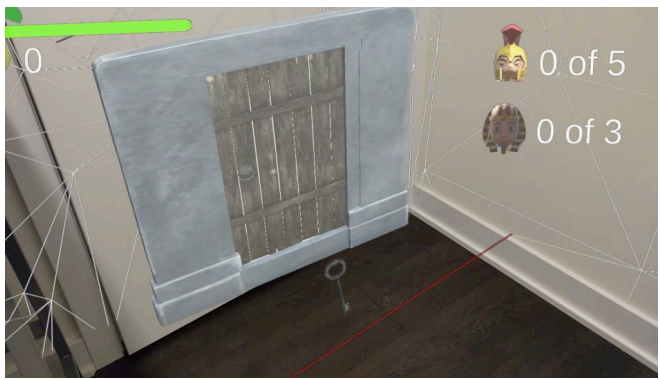

Fig 10. The door with the key to unlock it.

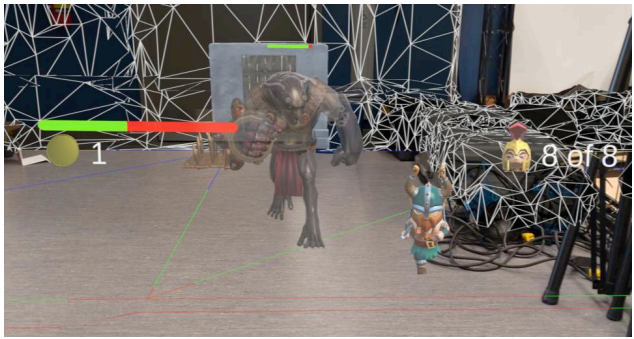

Fig 11. Boss fight in Level 2.

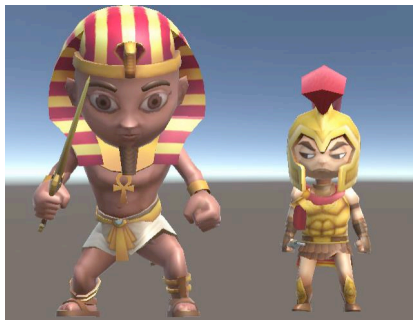

Fig 12. A showcase of mini-boss (Left) and mob (Right) in Level 3.

In the first level, players face mob (mobile) enemies and a unique boss, with 15 coins available for collection. Players must defeat five mobs to confront the boss. Defeating the boss unlocks a key, allowing access to the next level (see Fig 10). In the second level, players must beat eight mobs to challenge a distinct boss to get to the next level. There are 25 coins scattered around. A victory against the boss (Fig 11) presents a key, leading to the third level. In the third level, players fight mobs, mini bosses (Fig 12), and a unique final boss. 30 coins are available to gather. Players need to defeat five mobs and three mini-bosses to face the boss. Overcoming the boss lets players collect a key, marking their victory in the game.

### 4.2 First-Person Shooter (FPS) Game

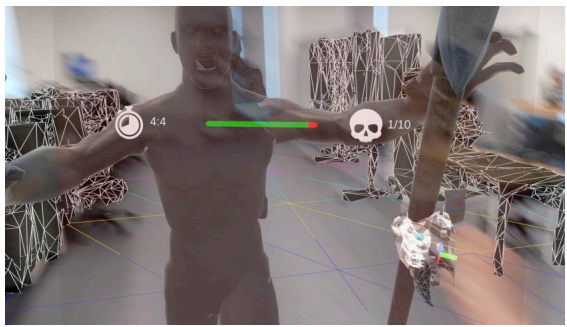

Fig 13. The utilization of the axe in Level 1.

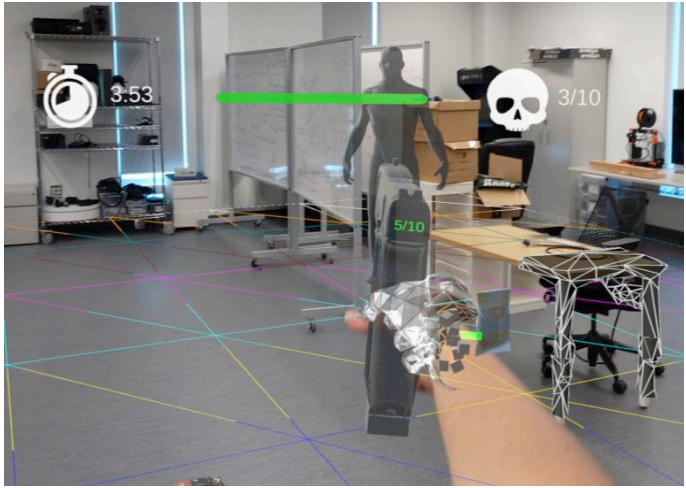

Fig 14. The utilization of the gun in Level 2.

The second game is an FPS set in a zombie survival scenario designed for the *HoloLens 2*, using its hand gestures and tracking capabilities. Players use hand movements to wield weapons and battle zombies. The aim is to eliminate a set number of zombies within a time frame, with the spatial mapping feature helping players strategize and navigate encounters. Each of the two levels introduces distinct gameplay mechanics.

In the first level, players use an axe to fend off 10 zombies (Fig 13). They activate the axe by a grabbing gesture and can collect virtual food to restore health points (HP). In the second level, players use a pistol to combat zombies (Fig 14). To hold the pistol, players need to maintain their fingers closed, except for their index and thumb fingers, which are used to grip the virtual gun. They fire by pinching. Ammo is limited, but pickups are available. Players must eliminate 10 zombies within 5 minutes to win.

### 4.3 Rationale for Developing These Two Games

Two games of different genres were developed to evaluate *GAMR*'s versatility across varied player experiences. This aimed to test *GAMR*'s adaptability to different genres, player experiences, perspectives, and inputs. The lead author meticulously polished the games based on feedback gathered through informal playtesting sessions and discussions with colleagues, ensuring all identified flaws and bugs were resolved before the *GAMR* evaluation. For instance, feedback indicated that the weapon trigger gesture in the second level of the FPS game, initially set to a curling point finger, was challenging due to *HoloLens 2*'s imperfect hand tracking. This gesture was changed to a pinch, which the headset detected more reliably. Additionally, in the Action-Adventure game, we were advised to add more audio feedback to enhance the player experience, as actions like dodging and jumping were not sufficiently clear. It was also recommended to add an indicator for when the player character was out of the field of view and difficult to navigate.

We developed an FPS game for combat focus and an Action-Adventure game for exploration. This diversity tests *GAMR*'s adaptability across varied game dynamics and scenarios. Using varied input methods like hand gestures in the FPS game tests *GAMR*'s adaptability. This evaluation helps gauge *GAMR*'s data handling from diverse inputs, offering insights for developers and designers.

## 5 Evaluation

We conducted a user study to assess *GAMR*'s usability and usefulness. The study focused on user interaction, ease of use, efficiency, and satisfaction, providing insights into its usability across two game genres. If *GAMR* simplifies the analysis process, uncovers challenging insights, or improves efficiency, it then can be deemed useful. The usefulness of *GAMR* depends on its ability to cater to the unique challenges of different game genres and provide valuable insights. For *GAMR* to be truly useful, it must align with the goals of game developers and designers.

The decision to develop games in these two genres was based on their widespread popularity worldwide [53]. To evaluate *GAMR*, the study was divided into two distinct phases. Phase 1 aimed to engage participants in the design and development of the games by playing them and providing feedback, as well as to collect data for the subsequent phase. Phase 2 focused on comparing *GAMR*'s performance across two distinct game genres, addressing research questions and offering insights into its utility in various gaming contexts. The primary research questions that guided this study were as follows:

1. For which game did users perceive *GAMR* to be most beneficial? And why?
2. What glitches or problems do users identify when using *GAMR*?
3. What aspects of *GAMR* do users appreciate or find unfavorable?
4. In what ways can *GAMR* be enhanced?
5. How does the actual experience with *GAMR* match up to users' initial assumptions?
6. Which other genres might find *GAMR* effective or ineffective? And why?

### 5.1 Phase 1: Collecting Data

The primary objectives of the initial phase were to familiarize users with game mechanics and gather essential data for the next phase. We also collected feedback on design and development to understand user approaches to game issues before using our analytic tool. This phase compared player experiences across two distinct games to highlight differences and provide insights for the second study. We integrated *GAMR* with the games to record gameplay data as participants played both games. In this phase participants played both games:

In this phase participants played both games:
- the FPS game consisting of two levels,
- the Action-Adventure game consisting of three levels.

*GAMR* was running in the background and recording as participants played the games.

#### 5.1.1 Participants

17 participants aged 18-33 (*Mdn* = 24) were recruited for this phase of the study: six females, nine males, one non-binary, and one undisclosed. We ran the study in our lab. Participants were picked from the demographic of undergraduate and graduate students in computer science and related disciplines at Ontario Tech University. On a pre-session form, five participants identified as "Hardcore Gamers," nine as "Core/Mid-Core Gamers," and three as casual gamers. 13 preferred both FPS and Action-Adventure games, three preferred Action-Adventure, and one preferred FPS games. Participants received a CAD 10 honorarium.

#### 5.1.2 Apparatus

Microsoft's *HoloLens 2* and *Xbox Series X controller* were the hardware that was used for the study. We also used Open Broadcaster Software (OBS) [54] for recording. Using the *Mixed Reality Capture* feature in the *HoloLens 2*'s *Windows Device Portal*, we streamed a live preview of the gameplay to a separate PC. *OBS* on this PC captured the stream, recording it as a video file, ensuring the *HoloLens 2*'s performance remained unaffected.

*5.1.3 Procedure*

Participants began with consent forms and a pre-session form, detailing demographics and gaming history. They were then briefly trained on the *HoloLens 2* and the games' objectives. During gameplay, *GAMR* tracked their actions and metrics. For a balanced evaluation, participants alternated between playing the Action-Adventure and FPS games. After completing each, they filled out the Player Experience Inventory (PXI) [1] questionnaire, offering insights into their gameplay experiences. At the end, we conducted a semi-structured interview with the participants to learn more about their experiences and to note any gameplay issues.

*5.1.4 Results and Discussion*

We analyzed the findings from using PXI responses and the semi-structured interview, which are presented below.

*5.1.4.1 Player Experience Inventory (PXI)*

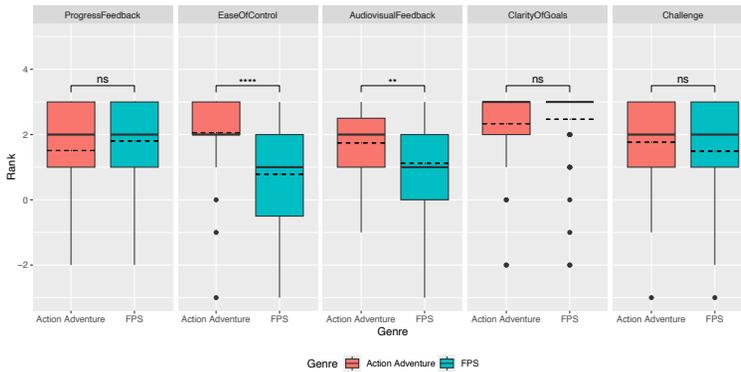

Fig 15. Functional consequences. A comparison between Action-Adventure and FPS using Wilcoxon signed-rank test. ns: p > .05, **: p < .01, ****: p < .0001.

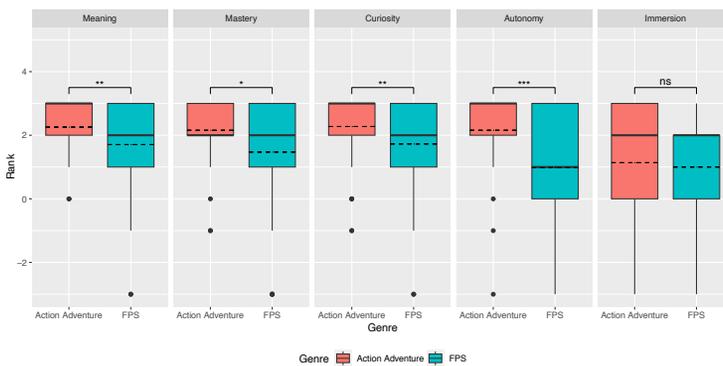

Fig 16. Psychosocial consequences. A comparison between Action-Adventure and FPS using Wilcoxon signed-rank test. ns: p > .05, *: p < .05, **: p < .01, ***: p < .001.

We conducted Wilcoxon tests with continuity correction and found significant differences in both functional and psychosocial consequences between the game genres. The results appear in Fig 15 and Fig 16. These findings highlight the importance of ease of control and audiovisual feedback in shaping the player experience and suggest that players may have a more positive experience with the Action-Adventure game in these aspects. These findings also suggest that the Action-Adventure game provided a greater sense of autonomy, curiosity, mastery, and meaning for players compared to the FPS game.

*5.1.4.2 Interview Findings*

After completing the study's initial phase, we interviewed participants to gather feedback on their gaming experience. Participants enjoyed the games, with P3 praising the action-adventure game's design and P4 the combat and 3D spatial mapping's immersion. Participants P05, P08, and P15 found the *HoloLens 2* headset's spatial mapping useful. However, feedback also pinpointed game issues. P08 and P17 mentioned double jump issues; P14, P13, and P07 found dodging tricky; and P03, P04, and P09 wanted more audio cues.

In the FPS game, participants liked the unique hand-gesture controls and spatial audio, which helped identify enemy positions. P14 emphasized the spatial audio's effectiveness, and P07 and P17 appreciated the game's visual clarity. P05 was notably immersed, stating their genuine startle while playing. Yet, challenges arose. P09 noted accidental gun firings, and P05, P08, and P13 experienced lag when many enemies appeared, a limitation of the *HoloLens 2*. Most participants found the axe mechanism smoother than the gun, indicating issues with the *HoloLens 2* recognizing pinching gestures. P12 faced zombies stuck in the spatial mapped area, affecting gameplay. Feedback also suggested enhancing gun aim for a better overall experience.

## 5.2 Phase 2: Evaluating *GAMR*

The second phase of the study focused on comparing the performance of *GAMR* across two distinct game genres. This phase aimed to address our research questions and provide insights into the usefulness of *GAMR* in different gaming contexts.

### 5.2.1 Participants

We recruited 20 participants aged 18-37 (*Mdn* = 24), comprising nine females, nine males, one non-binary, and one undisclosed gender. They received a $20 honorarium. Most participants were from the initial study phase, with three new additions, taken from the same demographic. A pre-session form gauged their experience on a 1-7 Likert scale. They were somewhat familiar with MR/AR (*Mdn* = 4) and VR games (*Mdn* = 5). Their game development and *Unity* experience was moderate (*Mdn* = 4 and 5 respectively). They felt competent in identifying game issues (*Mdn* = 5), had some VR game development experience (*Mdn* = 3), but lacked in MR/AR game development experience (*Mdn* = 1).

*5.2.1.1 Power Analysis*

This study, primarily exploratory due to the lack of prior research guiding sample size ([29], p. 171), used an *a priori* power analysis expecting a large effect size ($d \geq 0.8$) to estimate a rough target sample size. With $\alpha = 0.05$ and $1 - \beta = 0.9$, our target exceeds the minimum 15 samples for a 0.8 power, as per Field et al. [18] (see p. 58), but should be viewed as an initial estimate.

### 5.2.2 Apparatus

We used the same *HoloLens 2* headset and recorded sessions with *OBS* [54] on a separate PC. The live game session from the *HoloLens 2* was streamed via the *Windows Device Portal*, and *OBS* saved this as a video. This method ensured the *HoloLens 2*'s performance remained unaffected, providing participants an uninterrupted experience.

### 5.2.3 Procedure

Participants began each session with informed consent forms and a demographic survey. They received a 10-minute briefing on using the *GAMR* tool with *HoloLens 2*, followed by time for questions and exploring *GAMR*'s features. We addressed potential memory lapse by allowing participants to play each game level during sessions lasting 25-30 minutes. To minimize bias, participants evaluated game data from sessions other than their own.

Table 2. The order and number of log files loaded into *GAMR* for Action-Adventure game analysis.

| Participant # | Task 1 | Task 2 | Task 3 |
| --- | --- | --- | --- |
| P01, P04, P07, P10, P13, P16, P19 | 1 Player Data | 2 Players Data | 3 Players Data |
| P02, P05, P08, P11, P14, P17, P20 | 3 Players Data | 1 Player Data | 2 Players Data |
| P03, P06, P09, P12, P15, P18 | 2 Players Data | 3 Players Data | 1 Player Data |

Table 3. The order and number of log files loaded into *GAMR* in FPS game analysis.

| Participant # | Task 1 | Task 2 |
| --- | --- | --- |
| P01, P03, P05, P07, P09, P11, P13, P15, P17, P19 | 1 Player Data | 3 Players Data |
| P02, P04, P06, P08, P10, P12, P14, P16, P18, P20 | 3 Players Data | 1 Player Data |

The next step was the game analysis. Half the participants analyzed the Action-Adventure game first, while the other half began with the FPS game. Data from the initial 17 participants' play sessions were loaded into *GAMR* for analysis (See Table 2 and Table 3). Participants began each session with informed consent forms and a demographic survey. They received a 10-minute briefing on using the *GAMR* tool with *HoloLens 2*, followed by time for questions and exploring *GAMR*'s features. We addressed potential memory lapse by allowing participants to play each game level during sessions lasting 25-30 minutes. To minimize bias, participants evaluated game data from sessions other than their own.

We input various amounts of log files into *GAMR* for thorough coverage. This helped us evaluate the tool's functionality using both multi-data and individual data analyses. Due to *HoloLens 2*'s hardware limits, only three log files were loaded at once. Participants used *GAMR* to review each game level for 10-15 minutes but could finish earlier if they felt done. They were not given specific analysis guidelines, allowing them to use *GAMR* features as they saw fit. The tool recorded usage metrics throughout their analysis. After analyzing each level, we stopped the screen recording and participants filled out a questionnaire about *GAMR*'s features. They ranked these features based on usefulness, giving feedback on specific aspects and their overall evaluation of the tool in game analysis. After analyzing all game levels, participants completed a System Usability Scale (SUS) [6] questionnaire on *GAMR*'s usability, taking about 2 minutes. A subsequent interview discussed their views on *GAMR*'s usefulness, game rankings, favored and disliked features, and *GAMR*'s compatibility with other genres. Participants also shared all other noteworthy experiences from the session.

### 5.2.4  Results and Discussion

After the study, we statistically analyzed the quantitative data for significant patterns. The qualitative feedback from participants and interview responses were grouped by similar themes for a comprehensive analysis, ensuring a detailed review of both quantitative and qualitative results.

#### 5.2.4.1  Self-Developed Questionnaire

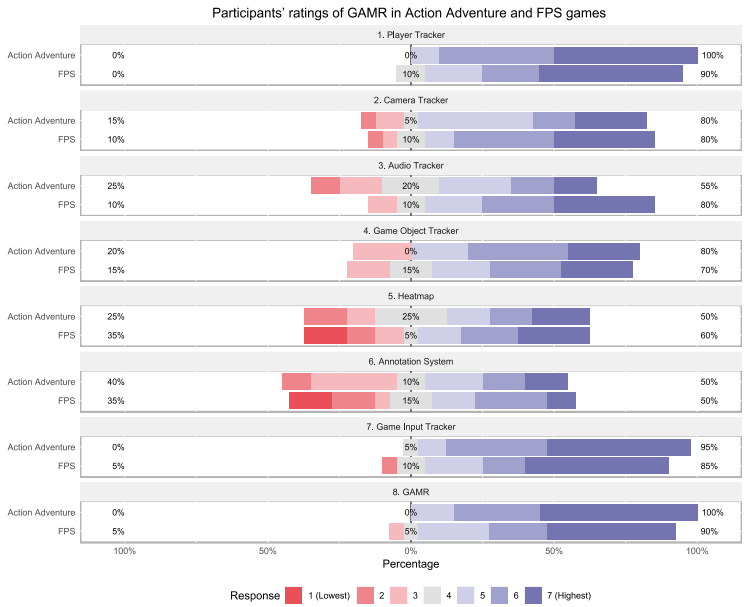

Fig 17. Summary of self-developed questionnaire rankings.

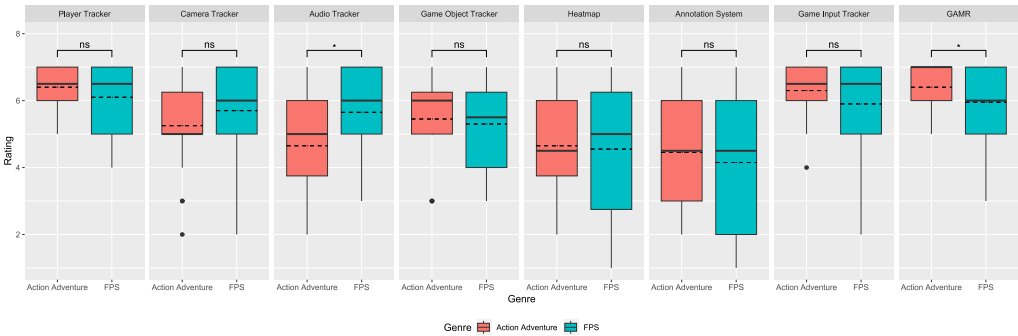

Fig 18. Summary of self-developed questionnaire's Wilcoxon signed-rank test. ns: p > .05, *: p < .05.

In the post-session questionnaire, participants ranked *GAMR* features' usefulness for each game genre on a 7-point Likert scale, from 1 (not useful) to 7 (very useful). The rankings for each feature and game genre appear in Fig 17.

The summarized findings appear in Fig 18. In most categories, there were no significant differences observed between the two game genres. However, notable distinctions were identified in the audio tracker and overall usefulness of *GAMR*. A Wilcoxon signed rank test revealed that the FPS game's audio tracker feature was rated higher ($Mdn = 6$) than the Action-Adventure game's ($Mdn = 5$), with a significant difference, $V = 20$, $p < .05$, $r = .323$. This is likely due to the FPS game's emphasis on audio, especially for understanding enemy spawn locations, which was not as vital in the Action-Adventure game. A Wilcoxon test also revealed that participants found *GAMR* more useful for the Action-Adventure game ($Mdn = 7$) than the FPS game ($Mdn = 6$), $V =$ 47, $p < .05$, $r = 0.176$. This indicates that *GAMR* may better support analysis in Action-Adventure games than in FPS games.

During the study, participants indicated their preferred number of log files and player data for *GAMR* analysis. For the Action-Adventure game, nine participants wanted three players' data, suggesting they valued analyzing multiple gameplay sessions. Meanwhile, three participants wanted two players' data, and eight wanted just one, showing varied preferences in data granularity for analyzing this game genre. For the FPS game, 11 participants preferred analyzing one player's data, showing an inclination towards individual analysis. Yet, nine participants wanted three players' data, highlighting an interest in comparing multiple gameplay sessions for this genre. These findings emphasize the need to consider individual preferences and data granularity when using *GAMR* for gameplay analysis. The varied preferences among participants underscore the importance of flexible data loading options to meet different analysis objectives and research goals.

*5.2.4.2 System Usability Scale (SUS)*

*GAMR* received a mean SUS score of 72.25 with SD = 11.94, $CI_{95\%}$ = [66.65, 77.84]. Being above 68, it is considered 'Acceptable'. According to the established interpretation resource, a SUS score in this range suggests that *GAMR*'s usability is above average, indicating that it was generally well-received by participants. Although participants generally responded positively, some faced challenges and required assistance, particularly regarding the learnability and ease of use of the system. This might be due to the unfamiliarity associated with using a novel system like this. Despite these challenges, none of the participants thought the system was unnecessarily complex.

Overall, the high ratings obtained from all participants across the game genres indicate that *GAMR* was perceived as usable. These positive ratings reflect participants' satisfaction with the usability of *GAMR* and highlight its effectiveness in supporting their game analysis tasks.

*5.2.4.3 Interview Findings*

At the study's conclusion, we held a semi-structured interview to gather participants' views on their *GAMR* experience. They were encouraged to share their thoughts and decision rationales. The 20 participants offered various insightful comments and perspectives, summarized below.

**Research Question 1 – "For which game did users perceive *GAMR* to be most beneficial? And why?"**

14 participants (P01-P08, P11-P15, P20) favored *GAMR* for the Action-Adventure game for several reasons:

a) The game presented a richer variety of trackable data, providing a more comprehensive session view.
b) The player tracker in the Action-Adventure game, displaying the 3D character's movement, was considered more useful.
c) For this game type, understanding intricate movements was essential, and *GAMR* facilitated this insight, helping game designers refine level design.
d) The Action-Adventure genre's intricate level designs made *GAMR*'s insights especially valuable.
e) The pace and dynamics of tracking in the Action-Adventure game meshed better with *GAMR*'s features.

However, six participants (P09, P10, P16-P19) felt *GAMR* was more suited for the FPS game because:

a)  The FPS game had fewer data elements, simplifying analysis.
b)  FPS games offered a more immersive experience, and *GAMR* effectively captured this immersion.
c)  The camera tracker feature was valued in the FPS game, providing a clear perspective on the player's field of view.

**Research Question 2 – "What glitches or problems do users identify when using *GAMR*?"**

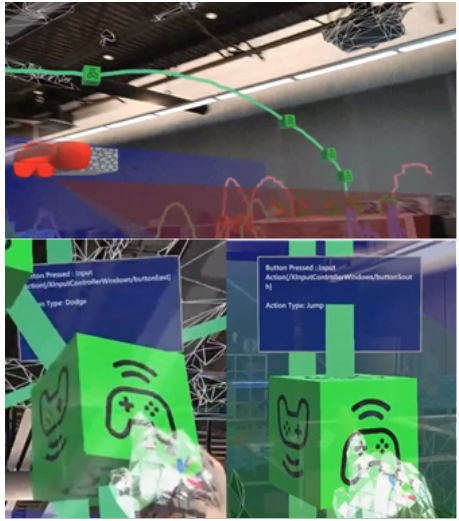

Fig 19. An issue identified by a participant: unusual player movement within the game area (top), and a button combination employed by the player that led to the issue, tracked via the game input monitor in *GAMR* (bottom).

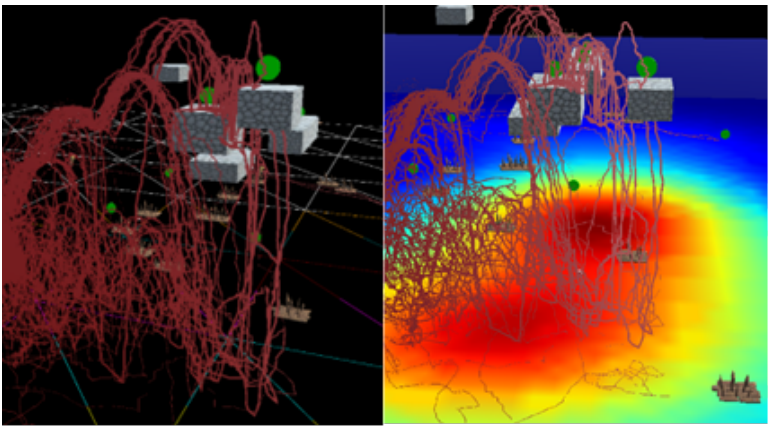

Fig 20. An issue linked to player experience and level design wherein players faced difficulties in gathering coins located on elevated platforms. A participant adeptly identified this problem utilizing both the replay system and heatmap.

Participants used *GAMR* to identify various game anomalies.
In the Action-Adventure game:

- P07 and P19 found an anomaly where a player used unconventional button combinations, resulting in unintended character movement (Fig 19).
- P02 and P04 noticed a character getting trapped behind a wall mapped by *HoloLens 2*.
- P18 felt the game objective lacked clarity.
- P04, P05, and P20 observed difficulties in coin collection, marking it as a level design flaw (Fig 20). They used *GAMR*'s replay and heatmap functionalities for this analysis.
- P09 noted players lingering at the first aid pack spawning point during boss battles.
- P09 also found a bug where the first aid pack got stuck above the ceiling, blocking progress.
- P15 and P20 identified enemies located outside the game space, making them hard to detect.

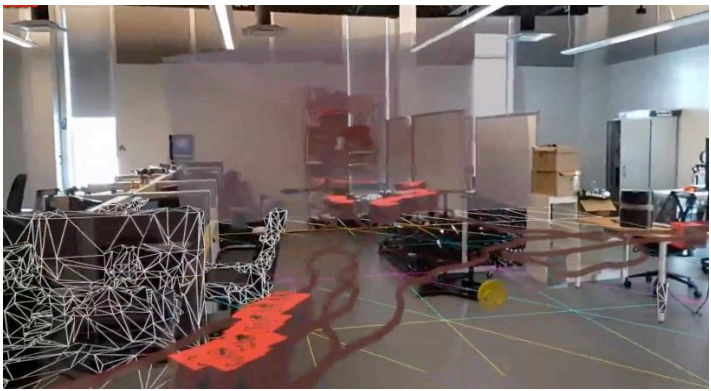

Fig 21. An identified problem in level design showing player movement constrained to a specific area due to insufficient randomness in enemy spawning.

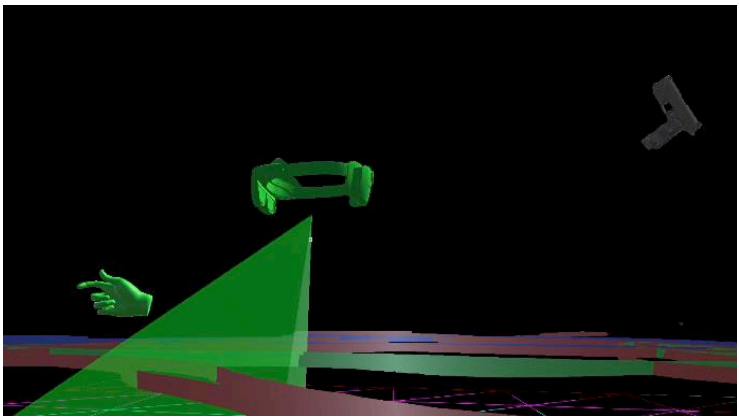

Fig 22. A bug identified: the gun game object becomes dislodged from the player's grip when they gaze downward or attempt to retrieve an in-game item from the floor.

In the FPS game:
- P05 spotted performance lags and inaccuracies in player movement tracking, particularly during fast turns, which caused hand tracking loss.

- P03 noticed enemies spawning predictably, making the game too easy.
- P11, P12, and P15 observed limited player movement across levels (Fig 21).
- P01, P07, P13, and P15 detected instances where the virtual gun was dislodged from players, causing game failure (Fig 22). This was linked to swift player movements.
- Using *GAMR*'s audio tracker, participants noted inconsistent audio feedback between both games. P03 and P11 felt the Action-Adventure game lacked sufficient audio feedback.
- With the camera tracker, participants identified unexplored game objects and optimal placement areas.
- P13 discovered a bug in the FPS game, which stopped players from collecting in-game items.

Overall, while the PXI results from phase 1 of the study provided valuable insights into initial player experiences and challenges, such as easier game controls and higher curiosity in the Action-Adventure game, the use of *GAMR* in phase 2 expanded upon these insights by offering a more detailed and data-driven approach to evaluating player interactions and gameplay mechanics.

While many issues identified with *GAMR* mirrored phase 1 findings, additional problems were also detected as stated above. Common issues included performance problems in the FPS game, such as a bug causing the gun to fly out of the player's hand, and the high effort required to collect certain coins in the Action-Adventure game, which reduced player motivation. Another issue was the boss enemy spawning out of bounds. These problems were confirmed by both participant feedback in phase 1 and *GAMR*'s game object tracker and playback system.

**Research Question 3 – "What aspects of *GAMR* do users appreciate or find unfavorable?"**

Participants universally praised *GAMR*'s player tracker feature for its ability to analyze multiple players' data, offering a comprehensive view of gameplay sessions.

The heatmap visualization garnered positive feedback from participants like P01, P03, P04, P06-P08, P12, P15-P20, proving essential for spotting common player patterns and level design issues.

Camera tracker was appreciated by P10, P14, P15, P17, and P20 for its insights into player perspectives. P01 and P03 valued the annotation system for real-time notetaking, while P04 and P13 praised the filter option for selective data analysis. P07 and P11 found the game input and audio trackers useful in the Action-Adventure game for detecting bugs and understanding audio elements, respectively.

P14 emphasized individual player data analysis's value, and P12 valued *GAMR*'s holistic data for understanding player interactions.

On the flip side, P02, P04, P11, P13, P14, and P16 found the overlap of loaded game data objects overwhelming. Some felt the heatmap was tricky to interpret, while P05, P08, and P12 deemed the annotation system unreliable compared to manual notetaking. P06 and P17 saw the audio tracker as insufficient, and P01 felt the camera tracker did not provide valuable insights. Lastly, P10 was not fond of viewing the entire recorded path at once.

These insights spotlight both the strengths and potential improvement areas of *GAMR*.

**Research Question 4 – "In what ways can *GAMR* be enhanced?"**

Participants provided suggestions for refining *GAMR*:

- **Player Tracking**: P04, P14, P17, and P18 requested improvements for simultaneous multi-player tracking and proposed a timestamp feature to avoid missing data.

- **Data Representation**: P07, P14, and P15 called for distinct shapes in the game controller and audio trackers, currently represented as 3D cubes, for better differentiation.
- **Heatmaps:** P04, P15, and P16 suggested the inclusion of 3D heatmaps for a more immersive understanding of gameplay.
- **Visualizing Overlapped Data:** P04 and P15 recommended dynamically adjusting the size of tracker objects based on data volume. P01 and P04 sought improvements for consistent data interaction and display.
- **Interaction:** P04 and P17 suggested adding game controller interaction for a more intuitive user experience. P15 and P19 proposed pointer interaction for high-placed data and toggling tracked objects' visibility.
- **Audio:** P10 and P04 felt audio commands would enhance functions like notetaking or playback control. P10 believed the audio tracker could be improved by auto-playing audio at the relevant replay point.
- **Indicators and Placement:** P01 sought indicators for player characters to avoid losing track during replays, while P13 requested accurate placement for the camera tracker and a feature to focus on specific game objects.
- **Color Coding:** P11 recommended color-coding recorded game objects for clear session associations.
- **Replay System**: P07 suggested a delay after note-saving before autoplaying to review notes.
- **Avatars:** P03 felt non-3D character players should have human-like avatars.

These recommendations highlight areas for potential enhancement in *GAMR*'s functionality and usability.

**Research Question 5 – "How does the actual experience with *GAMR* match up to users' initial assumptions?"**

Participants had certain unmet expectations from *GAMR*:

- **Health Status**: P04, P09, P15, and P20 sought information on the player and enemies' health to gauge gameplay dynamics.
- **Replay Interaction**: P04 wanted automatic pausing during interactions or rewinding to focused locations for easier navigation.
- **Headset Rotation & Animations**: P13, P16, and P20 expected the 3D headset models to mimic player camera rotation. P13 missed the presence of animations to capture actions more dynamically.
- **Object Placement**: P13 and P20 noticed some game objects were not precisely located, affecting data accuracy.
- **Color Mapping**: P11 proposed color differentiation for specific game objects to enrich gameplay analysis.

Despite these suggestions, most participants felt *GAMR* encompassed the essential features for game analytics.

**Research Question 6 – "Which other genres might find *GAMR* effective or ineffective? And why?"**

Participants shared varied expectations about the game genres *GAMR* would excel in:

- **Adventure, Puzzle-Adventure & Stealth**: P02-P07, P15, P16, and P18 believed these exploration-driven genres would benefit from *GAMR*'s player movement and interaction tracking.
- **Platformer, RPG & Open-world**: P01, P02, P06-P07, P11, P12, and P20 saw potential in *GAMR* for analyzing these genres, especially tracking player movement in real-world exploration.
- **VR, AR, & MR Games**: P05, P08, P13, P14, P15, P18, and P19 felt these immersive environments would align with *GAMR*'s tracking capabilities.
- **Racing Games**: P04, P09, and P15 believed the tool could highlight glitches in multi-player scenarios.
- **Survival-Horror Games**: P15 and P16 saw the tool as beneficial for examining player survival strategies.
- **Simulation Games**: P20 believed these decision-driven games would benefit from *GAMR*'s analytical capabilities.
- **Strategy Games**: P06 found potential in analyzing player movements and decision-making.
- **Games with Fewer Elements**: P10 thought the tool would excel with simpler games.

However, they also identified genres where *GAMR* might not be optimal:

- **Puzzle Games**: P03, P08, P09, P11, P15, P17, and P20 felt that the static environments of these games would not benefit much from movement tracking.
- **Fighting & Fast-Paced Games**: P02, P04, P11, P12, P16, and P19 thought the static nature of fighting games and the rapid movements in shooter games might not mesh well with *GAMR*. However, P02, P06, and P17 felt shooter games could still benefit.
- **MMO Games**: P03, P12, P18, and P19 cited the vast scale of these games as a potential challenge for *GAMR*.
- **Action Games**: P10 felt the tool may not be in line with the intense gameplay of these games.
- **Games with Linear Progression**: P07 believed *GAMR* might not offer much in games with a predetermined path and limited player choices.

This feedback underscores the importance of tailoring *GAMR* to the diverse demands of different game genres.

*5.2.4.4 GAMER's Usage Metrics*

Table 4. GAMR usage metrics.

| Metric | Description |
| --- | --- |
| NumTimesPlayed | Total number of "play" button presses |
| NumTimesPlayedReverse | Total number of "rewind" button presses |
| NumTimesPlayedForward | Total number of "fast forward" button" presses |
| NumTimesPaused | Total number of "pause" button presses |
| NumTimesheatmaoToggled | Total number of "heatmap" button presses (to toggle heatmap) |
| NumTimesNoteGenerated | Total number of the notes generated by the user |

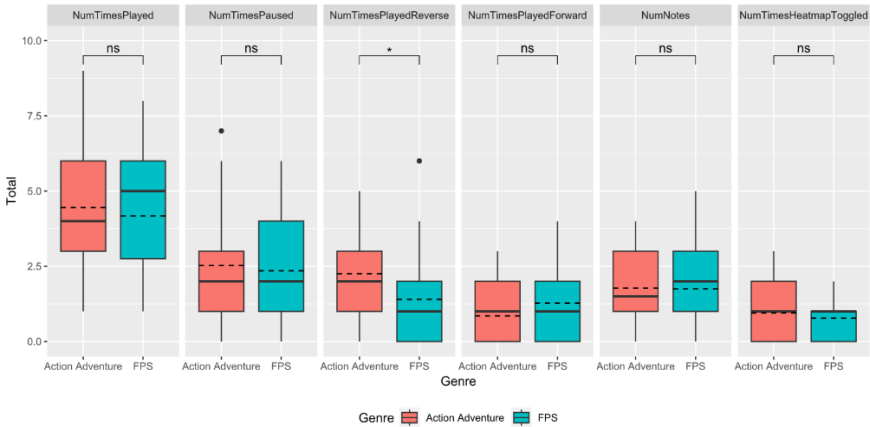

Fig 23. Summary of the recorded metrics' results. ns: p > .05, *: p < .05.

We collected usage metrics to understand participant interaction (Table 4). A Wilcoxon test with continuity correction revealed a significant difference in the total number of times participants pressed the "rewind" button (Fig 23), $V = 350.5$, $p < .05$, $r = .426$. More button presses were logged for the Action-Adventure game. These differences can be attributed to the larger amount of data analyzed in the Action-Adventure game compared to the FPS game. P03, P04, P15, and P18 stated that the Action-Adventure game had more interactable data, necessitating rewinding and pausing to analyze specific areas multiple times. However, the difference in pausing was not supported statistically at $\alpha = .05$, $V = 318$, $p < .1$, $r = .057$. Although no statistical differences were found in other metrics across the games, the findings indicate that all features were utilized, albeit with varying levels of popularity.

### 5.3 Overall Discussion

*GAMR* was developed with the aim of creating a versatile and seamless gameplay analysis tool specifically designed for MR games. By integrating features from game analytics and MR analytic tools, *GAMR* offers a distinctive solution. The primary objective was to develop a tool that accurately captures and presents gameplay data in a manner that closely resembles the original game session, while also granting users control over the recorded gameplay session.

The evaluation of *GAMR* conducted in this study yielded promising results, as participants responded positively to the tool across both the Action-Adventure and FPS game genres. Although the Action-Adventure game was deemed the most suitable genre for *GAMR*, participants acknowledged the tool's applicability to a wide range of game genres and its effectiveness in identifying bugs and issues in MR games. These findings highlight the potential of *GAMR* in diverse game development contexts. It is crucial to recognize the limitations identified in the study regarding *GAMR*. While *GAMR* may not completely replace traditional methods of analyzing MR games, such as video footage review, it presents a viable alternative that provides data comparable to video footage and extends beyond its capabilities. The results of the study suggest that *GAMR* has the potential to emerge as a valuable game analytics tool for MR game developers and designers.

By analyzing the PXI results and conducting interviews with players in phase 1 of the study, we gained insights into players' gameplay experiences, strengths, and weaknesses within the games. This data collection aimed to determine whether participants could employ *GAMR* effectively to discover the issues and bugs brought up by players in Phase 1, and whether they could extend beyond these known issues to identify new ones. The outcomes from Phase 2 indeed demonstrated that participants could leverage the diverse features of *GAMR* to uncover both previously identified issues and novel ones.

The outcomes and insights gleaned from the interviews conducted in phase 2 revealed that participants successfully identified bugs, which pertain to programming errors, flaws, or unintended behaviors within the game. Conversely, when it came to game-related issues centered around player experience, only a minority of participants managed to spot such concerns within the games. Consequently, it is evident that *GAMR* demonstrates a greater aptitude for detecting bugs within the games, thus making it a more fitting tool for aspiring game developers and designers. However, further comprehensive exploration is required to ascertain its full potential in the future.

The study has provided valuable insights into the strengths and areas for improvement of *GAMR*. It has shed light on how the tool can be enhanced and further developed in the future.

In conclusion, *GAMR* shows promise as a game analytics tool for MR games. With further improvements and refinements based on the study findings, it has the potential to become an asset for those seeking to analyze and optimize gameplay experiences in the MR domain.

## 6 Limitations

This study has notable limitations affecting its validity and generalizability. Firstly, the convenience sampling method relied on voluntary participation, potentially introducing biases and limiting the findings' generalizability beyond university computer science and game development programs. While these participants may not yet be fully specialized in MR game development, their background and training in game development provide a relevant and insightful perspective. The initial assumption of a large effect size for sample size determination may have been overly optimistic due to the absence of prior studies, suggesting that future research should consider larger sample sizes.

Secondly, technical limitations of the *HoloLens 2* headset, such as latency and field of view challenges, occasionally caused game and *GAMR* performance issues, potentially affecting participant experiences and subjective ratings. Using more advanced MR hardware in future studies could mitigate these concerns.

Thirdly, the study's games were relatively simple, and our findings are primarily based on the Action-Adventure and FPS genres due to restricted access to diverse open-source MR projects, potentially limiting the representation of varied game genres and impacting study outcomes. Employing more complex and diverse games in future evaluations could yield more comprehensive insights. For instance, adjustments in the analytic mechanisms might be necessary to accommodate the unique interaction patterns and design requirements of different game genres, such as puzzle games or simulations.

Lastly, the games were designed for a controlled university lab environment, which may have influenced participant experiences due to factors like distractions and varying equipment quality. Conducting evaluations in diverse real-world settings could provide a broader perspective on MR experiences. However, it should be noted that *GAMR* is currently not optimized for identifying issues related to dynamic changes in room layouts typical of home environments, as it was designed specifically around the layout of the Laboratory for Games and Media Entertainment Research (GaMER Lab) at Ontario Tech University.

## 7 Future Work

Insights from our study suggests several avenues for refining *GAMR*.

Addressing the lack of timestamps in the replay system, as pointed out by participants, is paramount. Future efforts should focus on enhancing the tool's data aggregation, given that game analytics tools are more effective with larger data volumes, revealing dominant patterns. Improving user interactions and making the interface more user-friendly will bolster the tool's usability. Participants also desired support for customizable objects in games, enabling developers to monitor specific elements in line with their game's mechanics.

Collaborating with real-world game projects can offer insights into *GAMR*'s practical applications and its potential influence on professional workflows. There's a need to streamline the tool's analysis process and delve into how various player behaviors are represented. While there's interest in enabling users to analyze dynamic game spaces adapting to player surroundings using *HoloLens 2*, the current hardware's limitations might impede optimal functionality.

MR devices offer a key advantage over traditional PC setups due to their portability, enabling users to work in various environments, allowing for context-aware analysis, where the game can adapt to the player's environment and situation. This can lead to a deeper understanding of how context influences gameplay and player experience. MR devices also facilitate collaborative efforts in a shared space, which is particularly beneficial for team-based analytics or educational settings, especially when the objective is to train individuals in conducting MR game analytics. These use cases are not currently supported by *GAMR*, but they represent potential areas for future development.

Video game analytics often neglect the player's perspective, despite growing interest in self-regulated learning and gameplay trends. Pfau and Seif El-Nasr [36] developed an analytics tool for *Guild Wars 2*, featuring encounter and log analysis, aligning with our *GAMR* tool's focus on boss encounter in the Action-Adventure game, and replays which align with both games. Wallner et al. [47] investigated players' needs in post-play visualizations across various genres. Players valued insights into decision impacts, resource efficiency, and understanding specific situations, with a focus on enemy positioning, playstyle, and item use. Visualization elements like movement, resource tracking, and fight analysis were deemed crucial. Our *GAMR* tool caters to these interests already. In the future, our work can be extended to help players inform their own behavior and decision-making in mixed reality games.

## 8 Conclusions

We developed and assessed *GAMR*, a tool tailored for bug detection in MR games for aspiring game developers. Post-development, we gathered data from 17 participants. Considering the disparities between game genres, we gathered data from two types: Action-Adventure and FPS. This differentiated approach pinpointed the unique elements of each genre. Another 20 participants then evaluated *GAMR*'s efficacy. Participants rated *GAMR* highly for usability across both genres, confirmed by SUS and our questionnaires. Semi-structured interviews further elucidated participants' positive feedback and insights. Despite some study limitations, the feedback affirmed *GAMR*'s utility for multi-genre game analysis.
In essence, *GAMR*'s development and assessment underscore its potential for MR game analysis. Its reception hints at its potential, and future iterations will address its shortcomings based on participant feedback.

## Acknowledgements

We used OpenAI's *ChatGPT-4* for editing and condensing the paper. However, it is important to note that the tool was not used in the creation of any original textual content.